\documentstyle[12pt]{article}

\catcode`\@=11 
\def\lesssim{\mathrel{\mathpalette\@versim<}}
\def\gtrsim{\mathrel{\mathpalette\@versim>}}
\def\@versim#1#2{\vcenter{\offinterlineskip
        \ialign{$\m@th#1\hfil##\hfil$\crcr#2\crcr\sim\crcr } }}
\catcode`\@=12 

\newcommand{\lw}[1]{\smash{\lower2.ex\hbox{#1}}}

   \def\cA{{\cal A}}
   \def\cD{{\cal D}}
   \def\drm{{\rm d}}

\begin{document}

\title{Renormalization Group Effects \\ on the Mass Relation Predicted \\
       by the Standard Model \\ with Generalized Covariant Derivatives}

\author{{\bf Tadatomi Shinohara, Kouzou Nishida,} \\
        {\bf Hajime Tanaka and Ikuo S. Sogami}    \\
        {\it Department of Physics, Kyoto Sangyo University, Kyoto 603}
       }

\date{\today}

\maketitle

\begin{abstract}
Renormalization group analysis is made on the relation
$m_{\rm H} \simeq \sqrt{2}m_t$ for masses of the top quark and the Higgs
boson, which is predicted by the standard model based on generalized
covariant derivatives with gauge and Higgs fields.
This relation is a low energy manifestation of a tree level constraint
which holds among the quartic Higgs self-coupling constant and
the Yukawa coupling constants at a certain high energy scale $\mu_0$.
With the renormalization group equation at one-loop level, the evolution
of the constraint is calculated from $\mu_0$ down to the low energy region
around the observed top quark mass. The result of analysis shows that
the Higgs boson mass is in $m_t \lesssim m_{\rm H} \lesssim \sqrt{2}m_t$
for a wide range of the energy scale $\mu_0 \gtrsim m_t$ and it approaches
to 177 GeV ($\approx m_t$) for large values of $\mu_0$.
\end{abstract}

\vskip 2.0cm
\begin{center}
Submitted to : {\it Progress of Theoretical Physics}
\end{center}

\newpage

  Confirmation of the existence of the top quark by the Collider Detector
Fermilab (CDF and D0) groups \cite{rf:1,rf:2} has shown that the standard
model is a consistent and correct theory of fundamental interactions.
To enrich the model further, however, we must solve many basic problems
remained concerning its scalar sector.  In particular, it is expected to
predict the Higgs boson mass for an experiment to observe it in near future.

  Recently, one of the authors \cite{rf:3,rf:4} has reformulated the standard
model by using the concept of generalized covariant derivatives with gauge
and Higgs fields which act on a multi-spinor field consisting of all
the chiral fermion fields. In the new model, the bosonic part of
the Lagrangian is determined by {\it field strengths} of the gauge and
Higgs fields which are constructed from the commutators of the generalized
covariant derivatives. As one of interesting outcomes of this scheme,
the strength of quartic self-interaction in the Higgs potential is fixed
exclusively in terms of the Yukawa coupling constants and, as a result,
the approximate relation $m_{\rm H} \approx \sqrt{2}m_t$ holds for the masses
of the top quark and the Higgs boson at the tree level.

  Unified description of the gauge and Higgs fields has been pioneered by
Connes~\cite{rf:5}.  Using the noncommutative geometry, he introduced
the Higgs field as a connection along the discrete direction in a doublely
sheeted Minkowski spacetime. His theory predicts the tree level relation
$m_{\rm top} = 2\,m_W$ and $m_{\rm Higgs} = 3.14\,m_W$.
{\'A}lvarez {\it et al}. \cite{rf:6,rf:7} investigated the evolution of
these relations under the one-loop renormalization group
equations~\cite{rf:8,rf:9}.  Following their method, we analyze quantum
effects on the restriction predicted by the standard model with
the generalized covariant derivatives in this article.

  Collecting three generations of the electroweak doublets $\psi_{lj}$ and
singlets $\psi_{ej}$ of lepton fields (the doublets $\psi_{qj}$, and singlets
$\psi_{uj}$ and $\psi_{dj}$ of quark fields) into a multi-spinor field,
we introduce the total fermion field
 \begin{equation}
   \Psi(x) = \sum_{j=1}^3 \sum_{\alpha = l,e,q,u,d}
             \psi_{\alpha j}(x) |\alpha j \rangle.
 \end{equation}
\noindent
The fermionic Lagrangian density is given by
 \begin{eqnarray}
    {\cal L}_f & = & i\,\sum_{j=1}^3 \sum_{\alpha = l,e,q,u,d}
    {\bar \psi}_{\alpha j} \gamma^\mu D_\mu \psi_{\alpha j}   \nonumber \\
       \noalign{\vskip 0.2cm}
               & &
  + \sum_{i,j=1}^3 \left\{
    a_{ij}^{(e)} {\bar \psi}_{li} \phi \psi_{ej}
  + a_{ij}^{(u)} {\bar \psi}_{qi} \tilde{\phi} \psi_{uj}
  + a_{ij}^{(d)} {\bar \psi}_{qi} \phi \psi_{dj}
  + {\rm h.c.}     \right\} ,
 \end{eqnarray}
where
 \begin{equation}
  D_\mu = \partial_\mu - ig_cA_\mu^{(3)a}{1 \over 2}\lambda_a
                       - igA_\mu^{(2)a}{1 \over 2}\tau_a
                       - ig'A_\mu^{(1)}{1 \over 2}Y
 \end{equation}
\noindent
is the ordinary covariant derivatives of the gauge group
${\rm SU(3)_c \times SU(2)_L}$ \linebreak
$\times {\rm U(1)}_Y$, $\phi$ is the Higgs doublet and
$a_{ij}^{(s)}\ (s = e,\,u,\,d)$ are the Yukawa coupling constants.
By factorizing ${\cal L}_f$ as
 \begin{equation}
   {\cal L}_f = \ :\bar{\Psi} i\gamma^\mu{\cal D}_\mu\Psi :\ 
              = -:{\rm Tr}\{ (i\gamma^\mu{\cal D}_\mu\Psi){\bar \Psi} \}:
 \end{equation}
we determine the generalized covariant derivative operator ${\cal D}_\mu$
acting upon $\Psi(x)$ in the form
 \begin{equation}
  {\cal D}_\mu = \partial_\mu - ig_c{\cal A}_\mu^{(3)}
                              - ig{\cal A}_\mu^{(2)} - ig'{\cal A}_\mu^{(1)}
               - {i \over 4} \gamma_\mu {\cal A}^{(0)}.
 \end{equation}
\noindent
Here $\cA_\mu^{(k)}\ (k=1,\,2,\,3)$ are the operator-valued gauge fields
defined by
 \begin{eqnarray}
  {\cal A}_\mu^{(3)} &=& A_\mu^{(3)a}{1 \over 2}\lambda_a
  \sum_{\alpha = q,u,d}|\alpha\rangle\langle\alpha|, \nonumber \\
  {\cal A}_\mu^{(2)} &=& A_\mu^{(2)a}{1 \over 2}\tau_a
  \sum_{\alpha = q,l}|\alpha\rangle\langle\alpha|, \nonumber \\
  {\cal A}_\mu^{(1)} &=& A_\mu^{(1)}{1 \over 2}
  \sum_{\alpha = l,e,q,u,d}y_\alpha|\alpha\rangle\langle\alpha|
 \end{eqnarray}
\noindent
with $y_l = -1$, $y_e = -2$, $y_q = 1/3$, $y_u = 4/3$ and $y_d = -2/3$, and
$\cA^{(0)}$ is the operator-valued Higgs fields
 \begin{eqnarray}
  {\cal A}^{(0)} & = &
  \sum_{ij}(\phi a_{ij}^{(e)} |{li}\rangle\langle{ej}|
           +\tilde{\phi} a_{ij}^{(u)} |{qi}\rangle\langle{uj}|
           +\phi a_{ij}^{(d)} |{qi}\rangle\langle{dj}| )  \nonumber \\
       \noalign{\vskip 0.2cm}
  & & +
  \sum_{ij}(\phi^\dagger a_{ij}^{(e)\ast} |{ej}\rangle\langle{li}|
           +\tilde{\phi}^\dagger a_{ij}^{(u)\ast} |{uj}\rangle\langle{qi}|
           +\phi^\dagger a_{ij}^{(d)\ast} |{dj}\rangle\langle{qi}| )
                                                          \nonumber \\
       \noalign{\vskip 0.2cm}
  & & +(c + c_5\gamma^5)
 \end{eqnarray}
\noindent
where $c$ and $c_5$ are real constants. For both the gauge and Higgs fields,
the field strengths ${\cal F}_{\mu\nu}^{(k)}$ $(k = 0,1,2,3)$ are introduced
into the theory through the commutator of the covariant derivatives as
 \begin{equation}
   [{\cal D}_\mu, {\cal D}_\nu] =
   - i \varrho_3 g_c {\cal F}_{\mu\nu}^{(3)}
   - i \varrho_2 g {\cal F}_{\mu\nu}^{(2)}
   - i \varrho_1 g' {\cal F}_{\mu\nu}^{(1)}
   - {i \over 4} \varrho_0 {\cal F}_{\mu\nu}^{(0)} ,
 \end{equation}
\noindent
where the factors $\varrho_k$ are specified by normalizing the kinetic
parts of the bosonic Lagrangian derived below. \par

The bosonic Lagrangian density ${\cal L}_b$ is determined by the field
strengths as
 \begin{eqnarray}
   {\cal L}_b &=& -{1 \over 4!}{1 \over 4}\sum_{k=0}^3{\rm Tr}
                 (\gamma^5 {\cal F}^{(k)}_{\mu\nu}
                  \gamma^5 {\cal F}^{(k)\mu\nu}) \nonumber \\
       \noalign{\vskip 0.2cm}
             & = &
  - {1 \over 4} \sum_{k=1}^3 F_{\mu\nu}^{(k)a} F_a^{(k)\mu\nu}
  + (D_\mu \phi)^\dagger (D^\mu \phi) + \mu^2 \phi^\dagger \phi
  - \lambda (\phi^\dagger \phi)^2.
  \label{bosonlag}
 \end{eqnarray}
\noindent
Namely, the bosonic part of the Lagrangian of the standard model is naturally
reproduced from the information of its fermionic part in this scheme.
The coefficient of the quartic term $(\phi^\dagger\phi)^2$ of
Eq.(\ref{bosonlag}) is, in definition, different from that of
Refs.~\cite{rf:7} and~\cite{rf:9} by the factor 2.  Furthermore, the constants
in the Higgs potential are determined as functions
of parameters appearing in the generalized covariant derivatives $\cD_\mu$ as
follows :
 \begin{equation}
  \mu^2 = c_5^2 - 3c^2
 \end{equation}
and
 \begin{equation}
  \lambda = \displaystyle{ {1 \over 2}
  {{\rm tr}\left\{(A_e^\dagger A_e)^2 + 3(A_u^\dagger A_u)^2
               + 3(A_d^\dagger A_d)^2 \right\} \over
   {\rm tr}\left\{(A_e^\dagger A_e) + 3(A_u^\dagger A_u)
               + 3(A_d^\dagger A_d) \right\}}
                         }, \quad
  A_s = (a_{ij}^{(s)}), \ \  s = e,u,d.
 \label{constraint}
 \end{equation}
\noindent
We interpret this relation between the quartic Higgs self-coupling $\lambda$
and the Yukawa coupling constants as the constraint at a certain high energy
scale $\mu_0$.

  It is straightforward to calculate the $\beta$ functions of the standard
model at the one-loop level~\cite{rf:9}.  By rewriting the ${\rm SU(3)_c}$,
${\rm SU(2)_L}$ and U(1)${}_Y$ gauge coupling constants ($g_c$, $g$ and
$g'$) as
 \begin{equation}
   g_3 = g_c,\ \  g_2 = g,\ \  g_1 = \displaystyle{\sqrt{5 \over 3}} g'
 \end{equation}
\noindent
and defining the $\beta$ functions by
 \begin{equation}
  \beta_i = \mu {\partial\alpha_i \over \partial\mu}, \quad
  \alpha_i = {g_i^2 \over 4\pi}
  \quad\quad (i = 1,2,3),
 \end{equation}
\noindent
we get
 \begin{equation}
  4\pi\beta_3 = -2\left({33 \over 3} - {4 \over 3}N_f\right)\alpha_3^2,
 \end{equation}
 \begin{equation}
  4\pi\beta_2 = -2\left({22 \over 3} - {4 \over 3}N_f
                                     - {1 \over 6}\right)\alpha_2^2,
 \end{equation}
 \begin{equation}
  4\pi\beta_1 = -2\left(- {4 \over 3}N_f - {1 \over 10}\right)\alpha_1^2,
 \end{equation}
\noindent
where $N_f$ is the number of fermion generations. Note the difference in
definitions of gauge coupling constants $g_3$, $g_2$ and $g_1$ in this article
and in Ref.~\cite{rf:3}. The matrix $A_u = (a_{ij}^{(u)})$ of the Yukawa
coupling constants for the up quark sector satisfies the evolution equation
 \begin{eqnarray}
  4\pi\mu{\partial A_u \over \partial\mu} & = &
   A_u \biggl[{1 \over 4\pi} {\rm tr} \left\{
       (A_e^\dagger A_e) + 3(A_u^\dagger A_u) + 3(A_d^\dagger A_d) \right\}
        \nonumber \\  \noalign{\vskip 0.2cm}
                                          & &
  + {3 \over 2} {1 \over 4\pi} \left( A_u^\dagger A_u - A_d^\dagger A_d \right)
  - {1 \over 2} \left(16\alpha_3 + {9 \over 2}\alpha_2
                                 + {17 \over 10}\alpha_1 \right) \biggr].
 \end{eqnarray}
\noindent
For the $\beta$ function of the quartic Higgs self-coupling $\lambda$
defined by
 \begin{equation}
  \beta_{\rm H} = \mu {\partial\alpha_{\rm H} \over \partial\mu}, \quad
  \alpha_{\rm H} = {\lambda \over 4\pi},
 \end{equation}
\noindent
we get
 \begin{eqnarray}
  4\pi\beta_{\rm H} & = & 24\alpha_{\rm H}^2
                  - {9 \over 5}\alpha_{\rm H}\alpha_1
                  - 9\alpha_{\rm H}\alpha_2 + {27 \over 200}\alpha_1^2
                  + {9 \over 20}\alpha_1\alpha_2 + {9 \over 8}\alpha_2^2
        \nonumber \\  \noalign{\vskip 0.2cm}
  & & + 4 {1 \over 16\pi^2}
          {\rm tr}\left\{(A_e^\dagger A_e) + 3(A_u^\dagger A_u)
                      + 3(A_d^\dagger A_d) \right\} \lambda
        \nonumber \\  \noalign{\vskip 0.2cm}
  & & - 2 {1 \over 16\pi^2}
          {\rm tr}\left\{(A_e^\dagger A_e)^2 + 3(A_u^\dagger A_u)^2
                      + 3(A_d^\dagger A_d)^2 \right\}.
 \end{eqnarray}
\noindent
In view of the large top quark mass in the low energy region, it is not
unnatural to assume that the Yukawa coupling constants are subject to
 \begin{equation}
   |a_{33}^{(u)}| \gg |a_{ij}^{(s)}|, \quad s = e,\,d\ ;\ (ij) \ne (33)
 \label{scale}
 \end{equation}
\noindent
for other energy scales also. This approximation simplifies
the renormalization group equations for $N_f = 3$ as
 \begin{equation}
  4\pi\beta_3 = -14\alpha_3^2
              = 4\pi{\drm \alpha_3 \over \drm t},
 \end{equation}
 \begin{equation}
  4\pi\beta_2 = -{19 \over 3}\alpha_2^2
              = 4\pi{\drm \alpha_2 \over \drm t} ,
 \end{equation}
 \begin{equation}
  4\pi\beta_1 =  {41 \over 5}\alpha_1^2
              = 4\pi{\drm \alpha_1 \over \drm t} ,
 \end{equation}
 \begin{equation}
  4\pi\beta_t = \alpha_t \left(9\alpha_t - 16\alpha_3 - {9 \over 2}\alpha_2
                                         - {17 \over 10}\alpha_1 \right)
              = 4\pi{\drm \alpha_t \over \drm t} ,
 \end{equation}
and
 \begin{equation}
  4\pi\beta_{\rm H}
              = 24\alpha_{\rm H}^2 + 12\alpha_{\rm H}\alpha_t
              - {9 \over 5}\alpha_{\rm H}\alpha_1 - 9\alpha_{\rm H}\alpha_2
              + {27 \over 200}\alpha_1^2 + {9 \over 20}\alpha_1\alpha_2
              + {9 \over 8}\alpha_2^2 - 6\alpha_t^2
 \label{rgehiggs}
 \end{equation}
\noindent
where $t = \ln \mu$, and the $\beta$ function for the Yukawa coupling constant
of the top quark was introduced by
 \begin{equation}
  \beta_t = \mu {\partial\alpha_t \over \partial\mu}, \quad
  \alpha_t = {(a_{33}^{(u)})^2 \over 4\pi} .
 \end{equation}
\noindent
Except for $\alpha_{\rm H}(t)$, these differential equations are analytically
solved as follows:
 \begin{equation}
  {1 \over \alpha_3(t)} - {1 \over \alpha_3(t_0)} =
  {7 \over 2\pi} (t - t_0) ,
 \label{gauge3}
 \end{equation}
 \begin{equation}
  {1 \over \alpha_2(t)} - {1 \over \alpha_2(t_0)} =
  {19 \over 12\pi} (t - t_0) ,
 \end{equation}
 \begin{equation}
  {1 \over \alpha_1(t)} - {1 \over \alpha_1(t_0)} =
  -{41 \over 20\pi} (t - t_0) ,
 \end{equation}
and
 \begin{equation}
  \alpha_t(t) = \displaystyle{
  {F_t^{213}(t) \over F_t^{213}(t_0)} \left[
  {\displaystyle{\alpha_t(t_0)} \over
   \displaystyle{1 - \alpha_t(t_0){9 \over 4\pi}
   \int_{t_0}^t {F_t^{213}(t) \over F_t^{213}(t_0)} \drm t}}
                                      \right] }.
 \label{alphat}
 \end{equation}
\noindent
In the last expression for $\alpha_t(t)$, the function $F_t^{213}(t)$ was
defined by \cite{rf:10}
 \begin{equation}
  F_t^{213}(t) = \alpha_3(t)^{8/7} \alpha_2(t)^{27/38}
                                   \alpha_1(t)^{-17/82}.
 \end{equation}

The scale relation in Eq.(\ref{scale}) for the Yukawa coupling constants
enables us to represent the masses of the Higgs boson and the top quark as
 \begin{equation}
  m_{\rm H}^2 = 2\mu^2 = 2\lambda v^2,\quad
  m_t \approx a_{33}^{(u)} {v \over \sqrt{2}}
 \end{equation}
\noindent
in terms of the vacuum expectation value $v = \sqrt{2}\langle \phi^0\rangle$
and to approximate the constraint in Eq.(\ref{constraint}) by
 \begin{equation}
 \lambda = {1 \over 2}\left( a_{33}^{(u)2} \right).
 \label{appconst}
 \end{equation}
\noindent
In order to investigate the deviation of this relation due to quantum effects,
we impose the initial condition
 \begin{equation}
  \alpha_{\rm H}(\mu_0) \approx {1 \over 2} \alpha_t(\mu_0)
 \label{initial}
 \end{equation}
\noindent
at a high energy scale $\mu = \mu_0$ and calculate the Higgs boson mass
$m_{\rm H}(\mu)$ at a lower energy scale $\mu$ by solving the renormalization
group equation. It is the tree level approximation that Eq.(\ref{appconst})
leads to the mass formula
 \begin{equation}
  m_{\rm H} \approx \sqrt{2} m_t.
 \label{tree}
 \end{equation}

The CDF collaboration \cite{rf:1} and the D0 collaboration \cite{rf:2} have
reported the value of the top quark to be
 \begin{equation}
   m_t = 180 \pm 12 ({\rm stat})^{+19}_{-15} ({\rm syst}) \  {\rm GeV}
 \end{equation}
\noindent
from the data of $\,p\bar{p}$ collisions at $\sqrt{s} = 1.8$ TeV.
As the first step to solve the renormalization group equation, let us
adopt the value $m_t = 180$ GeV for the top quark mass and decide
the values of the input parameters at the scale $\mu = m_Z = 91.2$ GeV. The
gauge coupling constants take the values
 \begin{equation}
  \alpha_3(\mu = m_Z) = 0.12,  \quad
  \alpha_2(\mu = m_Z) = 0.034, \quad
  \alpha_1(\mu = m_Z) = 0.017,
 \label{inigauge}
 \end{equation}
\noindent
and the vacuum expectation value is estimated to be
 \begin{equation}
  {\bar v} = v(m_Z)
           = {M_Z(m_Z)\cos\theta_W(m_Z) \over
              [\pi\alpha_2(m_Z)]^{1/2}} 
           = (\sqrt{2} G_F)^{-1/2}
           = 246\,{\rm GeV}.
 \end{equation}
\noindent
Integrating the renormalization group equations from
$m_t = m_t(m_t) = 180 \,{\rm GeV}$ to $m_Z$, we get
 \begin{equation}
    m_t(m_Z) \approx a_{33}^{(u)}(m_Z) {{\bar v} \over \sqrt{2}}
    = 187.9 \,{\rm GeV} ,  \qquad
    \alpha_t(m_Z) = 0.09283 .
   \label{inialphat}
 \end{equation}

\begin{figure}
\setlength{\unitlength}{1mm}
\begin{picture}(80,30)
 \put(35,0){\framebox(70,30){ }}
\end{picture}
\caption{Running behaviors of the Yukawa coupling constant
squared $\alpha_t(\mu) = (a_{33}^{(u)}(\mu))^2/4\pi$ and the quartic Higgs
self-coupling constant $\alpha_{\rm H}(\mu) = \lambda(\mu)/4\pi$ versus
the scale $\mu$.  The thick solid line representing the running of
$\alpha_t(\mu)$ is uniquely fixed by the experimental condition
$m_t(180\ {\rm GeV}) = 180$ GeV.  The dashed line is for $\alpha_t(\mu)/2$.
The thin solid lines showing the running of $\alpha_{\rm H}(\mu)$ have
dependence on the initial scale $\mu_0$.
The constraint $\alpha_{\rm H} = \alpha_t /2$ at an initial scale $\mu_0$
determines the evolution curve $\alpha_{\rm H}(\mu)$.
Three evolution curves $\alpha_{\rm H}(\mu)$ are drawn for the initial
scales $\mu_0 = 10^4$, $10^8$ and $10^{12}$ GeV.}
\label{fig:1}
\end{figure}
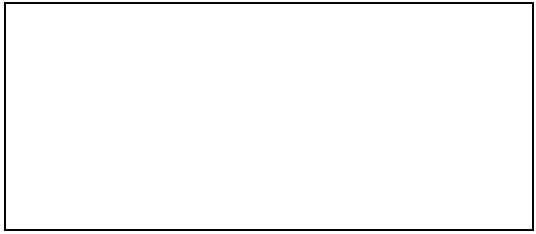

  Substitution of the input values in 
Eqs.(\ref{inigauge})$\sim$(\ref{inialphat}) into
Eqs.(\ref{gauge3})$\sim$(\ref{alphat}) determines uniquely the evolutions of
$\alpha_k\ (k=1,\,2,\,3)$ and $\alpha_t$.  Fig.~\ref{fig:1} shows
the evolution curve of $\alpha_t(\mu)$, which fixes the value of
$\alpha_{\rm H}(\mu_0)$ by the condition in Eq.(\ref{initial})
for each $\mu_0$.  Then, with such initial values for $\alpha_{\rm H}(\mu_0)$,
the renormalization group equation (\ref{rgehiggs}) is numerically solved.
In Fig.~\ref{fig:1}, the behaviors of the $\alpha_{\rm H}(\mu)$ curves are
plotted for $\mu_0 = 10^4$, $10^8$ and $10^{12}$ GeV.

We estimate here the running effect of the vacuum expectation value $v$. The
renormalization group equation for $v$ is obtained \cite{rf:9} by
 \begin{equation}
  {\drm \ln v \over \drm t} = {1 \over 16\pi^2} \left[
  {9 \over 4}\left({1 \over 5}g_1^2 + g_2^2\right) -
  {\rm tr}\left\{(A_e^\dagger A_e) + 3(A_u^\dagger A_u)
                                   + 3(A_d^\dagger A_d) \right\}
                                                \right] .
 \end{equation}
\noindent
Using Eq.(\ref{scale}), this is approximated by
 \begin{equation}
  4\pi{\drm v \over \drm t} =
  -v \left(3\alpha_t - {9 \over 4}\alpha_2 - {9 \over 20}\alpha_1 \right)
 \end{equation}
\noindent
which has the solution
 \begin{equation}
  v(t) = v(t_0) \left({\alpha_t(t) \over \alpha_t(t_0)}\right)^{-1/3}
         \left({\alpha_3(t) \over \alpha_3(t_0)}\right)^{8/21}
         \left({\alpha_2(t) \over \alpha_2(t_0)}\right)^{-9/76}
         \left({\alpha_1(t) \over \alpha_1(t_0)}\right)^{-7/492} .
 \end{equation}
\noindent
This running effect is confirmed numerically to be negligible \cite{rf:9} for
the range of mass scale considered here. Therefore, we use the following
formulas as
 \begin{equation}
  m_t(\mu) =
  \sqrt{2\pi\alpha_t(\mu)} \: v(\mu) \simeq
  \sqrt{2\pi\alpha_t(\mu)} \: {\bar v}
 \end{equation}
and
 \begin{equation}
  m_{\rm H}(\mu) =
  \sqrt{8\pi\alpha_{\rm H}(\mu)} \: v(\mu) \simeq
  \sqrt{8\pi\alpha_{\rm H}(\mu)} \: {\bar v}
 \end{equation}
for the running masses of the top quark and the Higgs boson for an arbitrary
scale $\mu$. \par

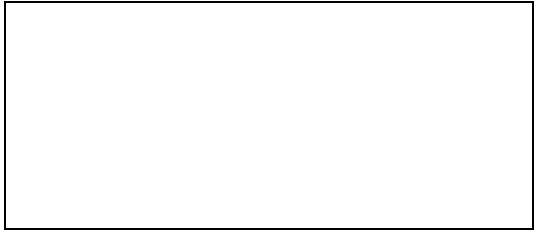
\begin{figure}
\setlength{\unitlength}{1mm}
\begin{picture}(80,30)
 \put(35,0){\framebox(70,30){ }}
\end{picture}
\caption{Dependence of the Higgs boson mass on the initial
scale $\mu_0$. The dashed line represents the $\mu_0$ dependence of the
Higgs mass $m_{\rm H}(\mu)$ at the scale $\mu = m_t = 180{\rm GeV}$.
Likewise the solid line is for the Higgs mass defined by the condition
$m_{\rm H}(\mu = m_{\rm H}) = m_{\rm H}$.
The relation $m_t \leq m_{\rm H}(m_{\rm H}),\, m_{\rm H}(m_t)
\leq 2^{1/2}m_t$ holds for a wide range of the scale $\mu_0 \geq m_t$.
With the increase of the scale $\mu_0$, the Higgs mass approaches to
177 GeV ($\approx m_t$).}
\label{fig:2}
\end{figure}

  Fig.~\ref{fig:2} represents the dependence of the Higgs boson mass on
the initial scale $\mu_0$. The dashed line is for the Higgs mass
$m_{\rm H}(\mu)$ at the scale $\mu = m_t = 180$ GeV. The solid line is
for the Higgs mass defined by the condition
 \begin{equation}
   m_{\rm H}(\mu = m_{\rm H}) = m_{\rm H}.
 \end{equation}
\noindent
The shapes of both lines are almost the same with each other and
both lines tend to converge to the common value $m_t = 180$ GeV
as the scale $\mu_0$ increases. From this result we find that the relation
$m_t \lesssim m_{\rm H}(m_{\rm H}), \,m_{\rm H}(m_t) \lesssim \sqrt{2}m_t$
holds for a wide range of the scale $\mu_0 \gtrsim m_t$ and that,
with the increase of the scale $\mu_0$, the Higgs mass approaches
to 177 GeV ($\approx m_t$).\footnote{Note that the masses $m_{\rm H}$
and $m_t$ are not physical pole masses. In Ref.~\cite{rf:11},
the relationship between the running mass and the physical pole mass is given
and the difference between them is proved to be negligible.}  \par

 In the analysis of {\'A}lvarez {\it et al}. \cite{rf:7} which has two
constraints $m_{\rm top} = 2\,m_W$, $m_{\rm Higgs} = 3.14\,m_W$ as initial
conditions for the renormalization group equation, both the value of
the Higgs boson mass and the value of the initial mass scale $\mu_0$ are
determined at the same time.  For example if we use the value $m_t = 186$ GeV
as the top quark mass at the scale $m_Z$, we get $\mu_0 \sim 10^4$ GeV and
$m_{\rm H} \sim 223$ GeV from Table 1 in Ref.~\cite{rf:7}.
By contrast we have only one constraint in Eq.(\ref{initial}) in our model.
This means that, even when the value of the top quark mass is given,
$\mu_0$ remains as free parameter.  However, it is natural to assume that
the scale $\mu_0$ at which the relation in Eq.(\ref{initial}) holds in
the original Lagrangian density for local fields takes a sufficiently
large value.  Therefore the results of our analysis show that the Higgs
boson has the mass being close to that of the top quark, {\it i.e.},
$m_{\rm H} \approx m_t$.

Experiments at LEP exclude a large range of Higgs masses.  Currently,
the LEP precision tests fixed the lower bound to be $m_{\rm H} = 58.4$ GeV
at 95\% confidence level~\cite{rf:12}.  On the other hand, a theoretical
constraint of the Higgs mass can be obtained from the vacuum stability
requirement that our universe is in the true minimum of the Higgs
potential~\cite{rf:13}.  The constraint depends upon the top quark mass and
upon the scale $\Lambda$ up to which the Standard Model remains valid.
In case where the constraint is severest, {\it i.e.}, $\Lambda = 10^{19}$ GeV,
$m_{\rm H} > 135\ {\rm GeV} + 2.1 (m_{\rm top} - 174
\ {\rm GeV})$~\cite{rf:14,rf:15}.  By non-perturbative calculations using
lattice field theory, an upper bound on the Higgs mass is obtained as
$m_{\rm H} < 710 \pm 60$ GeV~\cite{rf:16}.  Thus our predictions of
the Higgs mass obtained in this letter are within the allowed bound for
both experiment and theory.\par

Eq.(\ref{rgehiggs}) shows that the top quark Yukawa coupling constant
$a_{33}^{(u)}$ gives a negative contribution to the $\beta$ function
$\beta_{\rm H}$ of the Higgs self-coupling $\lambda$. Owing to the large top
quark mass, the value of $\beta_{\rm H}$ is always negative at low energy
scale. Since the top quark Yukawa coupling constant itself decreases with
scale, the value of $\beta_{\rm H}$ at high energy scale is positive.
Eventually, $\lambda$ falls with scale until some minimization is
reached, and then rise.  If this minimum is above zero, the standard model
vacuum is stable~\cite{rf:13,rf:14,rf:17}.  As shown in Fig.~\ref{fig:1},
the minimum of $\lambda$ is positive for all values of $\mu_0$ under our
initial condition (\ref{initial}).  Therefore, the vacuum is always stable
in our model.

After $\lambda$ reaches to some minimization, the $\beta$ function of
the Higgs self-coupling $\beta_{\rm H}$ is positive, and thus the Higgs
self-coupling $\lambda$ will eventually diverge, reaching to the Landau pole
(the Landau ghost)~\cite{rf:18}.  Once $\lambda$ exceeds unity it will diverge
rapidly. 
\begin{table}
 \caption{$\mu_0$ dependence of the Landau pole.}
 \label{table:1}
 \begin{center}
 \begin{tabular}{c|c}  \hline\hline
   $\mu_0$ & Scale reaching to the Landau pole \\ \hline
      500    & $2.5 \times 10^8$~GeV     \\ \hline
    $10^4$   & $1.7 \times 10^{11}$~GeV  \\ \hline
    $10^6$   & $3.1 \times 10^{15}$~GeV  \\ \hline
    $10^8$   & $5.5 \times 10^{19}$~GeV  \\ \hline
   $10^{10}$ & $1.1 \times 10^{24}$~GeV  \\ \hline
   $10^{12}$ & $2.5 \times 10^{28}$~GeV  \\ \hline
   $10^{14}$ & $9.2 \times 10^{32}$~GeV  \\ \hline
   $10^{16}$ & $7.8 \times 10^{37}$~GeV  \\ \hline
 \end{tabular}
 \end{center}
\end{table}
In other words there is no practical difference between the scale
where $\lambda$ tends to 
diverge and the non-perturbative scale corresponding
to $\lambda \geq 1$. In Table~\ref{table:1} we show the scale where
$\lambda$ diverges for each $\mu_0$. The scale giving a Landau pole is not
directly proportional to $\mu_0$.  It increases approximately as $\mu_0^{2.2}$
as $\mu_0$ increases. This means that the large value of $\mu_0$ leads to
the small masses of the top quark and the Higgs boson, so that the increase
of $\lambda$ becomes slow.  For low energy scale, $a_{33}^{(u)}$ and
$\lambda$ take finite values.\par

In this way we have analyzed the effects of one loop quantum corrections on
the constraint among the Yukawa coupling constants and the quartic Higgs
self-coupling constant predicted by the new scheme of the standard model
based on the generalized covariant derivatives. Numerical analysis of the
renormalization group equation has shown that the masses of the Higgs boson
and the top quark satisfy the relation $m_{\rm H} \approx m_t$ which deviates
markedly from the tree level prediction $m_{\rm H} \approx \sqrt{2}m_t$.
It is necessary to investigate the effects of quantum corrections on various
constraints among coupling constants which are obtained in the grand unified
theory \cite{rf:19} based on the generalized covariant derivatives.


\begin{thebibliography}{99}
 \bibitem{rf:1}
    CDF Collaboration, F. Abe {\it et al}., Phys. Rev. Lett.
                      {\bf 74}, 2626 (1995); {\bf 75}, 3997 (1995).
 \bibitem{rf:2}
    D0 Collaboration, S. Abachi {\it et al}.,
                                Phys. Rev. Lett. {\bf 74}, 2632 (1995);
                                Phys. Rev. D {\bf 52}, 4877 (1995).
 \bibitem{rf:3}
    I.~S.~Sogami, Prog. Theor. Phys. {\bf 94}, 117 (1995).
 \bibitem{rf:4}
    K.~Morita, Prog. Theor. Phys. {\bf 94}, 125 (1995);
    K.~Morita, Y.~Okumura and M.~T.-Yamawaki, {\it ibid}.
                                              {\bf 94}, 445 (1995);
    K.~Morita and Y.~Okumura, {\it ibid}. {\bf 95}, 227 (1996).
 \bibitem{rf:5}
    A.~Connes, Noncommutative Geometry (Academic Press, London, 1994);
    A.~Connes and J.~Lott, Nucl. Phys. {\bf B} (Proc. Suppl.)
                           {\bf 18}, 29 (1990);
    A.~H.~Chamseddine, G.~Felder and J.~Fr{\"o}lich,
                           Nucl. Phys. {\bf B353}, 689 (1991).
 \bibitem{rf:6}
    E.~{\'A}lvarez, J.~M.~Gracia-Bond{\'\i}a and C.~P.~Mart{\'\i}n,
                          Phys. Lett. B {\bf 306}, 55 (1993).
 \bibitem{rf:7}
    E.~{\'A}lvarez, J.~M.~Gracia-Bond{\'\i}a and C.~P.~Mart{\'\i}n,
                          Phys. Lett. B {\bf 329}, 259 (1994).
 \bibitem{rf:8}
    J. Collins, Renormalization (Cambridge University Press,
       Cambridge, 1984).
 \bibitem{rf:9}
    H.~Arason {\it et al}., Phys. Rev. D {\bf 46}, 3945 (1992);
    M.~E.~Machacek and M.~T.~Vaughn, Phys. Lett. {\bf 103B}, 427 (1981);
          Nucl. Phys. {\bf B222}, 83 (1983); {\bf B236}, 221 (1984);
                      {\bf B249}, 70 (1985).
 \bibitem{rf:10}
    G.~Jungman, Phys. Rev. D {\bf 46}, 4004 (1992).
 \bibitem{rf:11}
    J.~A.~Casas, J.~R.~Espinosa, M.~Quir{\'o}s and A.~Riotto,
       Nucl. Phys. {\bf B436}, 3 (1995);
       Erratum:  {\it ibid}. {\bf B439}, 466 (1995).
 \bibitem{rf:12}
    Particle Data Group, L.~Montanet {\it et al}.,
       Phys. Rev. D {\bf 50}, 1173 (1994)
       and 1995 off-year partial update for the 1996 edition
       available on the PDG WWW pages (URL: http://pdg.lbl.gov/).
 \bibitem{rf:13}
    M.~Sher, Phys. Rep. {\bf 179}, 273 (1989).
 \bibitem{rf:14}
    G.~Altarelli and G.~Isidori, Phys. Lett. B {\bf 337}, 141 (1994).
 \bibitem{rf:15}
    J.~R.~Espinosa and M.~Quir{\'o}s, Phys. Lett. B {\bf 353}, 257 (1995).
 \bibitem{rf:16}
    U.~M.~Heller, M.~Klomfass, H.~Neuberger and P.~Vranas,
       Nucl. Phys. {\bf B405}, 555 (1993).
 \bibitem{rf:17}
    M.~Sher, Phys. Lett. B {\bf 317}, 159 (1993);
       Addendum: {\it ibid}. {\bf 331}, 448 (1994).
 \bibitem{rf:18}
    L.~D.~Landau and I.~Pomeranchuk, Dokl. Akad. Nauk. USSR {\bf 102},
       489 (1955);
    L.~D.~Landau, in {\it Niels Bohr and the Development Physics}, edited
       by W.~Pauli (Pergamon, London, 1955).
 \bibitem{rf:19}
    I.~S.~Sogami, Prog. Theor. Phys. {\bf 95}, 637 (1996).
\end{thebibliography}
\end{document}